\documentclass[12pt]{iopart}
\usepackage[dvips]{graphicx} 
\usepackage{amssymb}
\usepackage{epsfig}
\usepackage{color}
\usepackage{ifthen}
\usepackage{iopams}  
\usepackage{multirow} 	
\begin{document}
\def\p{\partial}
\def\half{{1\over 2}}
\def\({\left(}
\def\){\right)}
\def\[{\left[}
\def\]{\right]}
\def\be{\begin{equation}}
\def\ee{\end{equation}}
\def\beq{\begin{eqnarray}}
\def\eeq{\end{eqnarray}}

\title[Residual foreground contamination in the WMAP data...]{Residual
  foreground contamination in the WMAP data and bias in
  non-Gaussianity estimation} 
\author{Pravabati Chingangbam$^{1,2}$} 
\ead{prava@iiap.res.in} 
\author{Changbom Park$^{2}$} 
\ead{cbp@kias.re.kr} 
\address{$^1$ Indian Institute of Astrophysics, Koramangala II Block,
  Bangalore  560 034, India\\ 
and \\
$^2$Korea Institute for Advanced Studies, 85 Hoegiro, Dongdaemun-gu,
Seoul 130-722, Korea} 

\begin{abstract}
We analyze whether there is any residual foreground contamination in
the cleaned WMAP 7 years data for the differential
assemblies, Q, V and W. We calculate the correlation between the
foreground map, from which long wavelength correlations have been
subtracted, and the foreground reduced map for each differential
assembly after applying 
the Galaxy and point sources masks. We find positive correlations for
$all$ the differential assemblies, with high statistical significance.
For Q and V, we find that a large fraction of the
contamination comes from
pixels where the foreground maps have positive values larger than
three times the rms values. These findings imply the presence of
residual contamination from Galactic emissions and unresolved point
sources. We redo the analysis after masking the extended point sources
cataloque of Scodeller {\em et al.}~\cite{Scodeller:2012fi} and 
find a drop in the correlation and corresponding significance
values. To quantify the effect of the residual
contamination on the search for primordial non-Gaussianity in the
CMB we add estimated contaminant fraction to simulated
Gaussian CMB maps and calculate the characteristic non-Gaussian
deviation shapes of Minkowski Functionals that arise due to the
contamination. We find remarkable agreement of these deviation shapes
with those measured from WMAP data, which imply that a major fraction
of the observed non-Gaussian deviation comes from residual foreground
contamination. We also compute non-Gaussian deviations of Minkowski
Functionals after applying the point sources mask of Scodeller {\em et
  al} and find a decrease in the overall amplitudes of the deviations
which is consistent with a decrease in the level of 
contamination.

\end{abstract}
\maketitle

\section{Introduction} 

The cosmic microwave background (CMB) radiation and the large scale 
structures in the universe carry a wealth of
cosmological information. Observational data support the
cosmological models dominated by cold dark matter and the cosmological
constant~\cite{Komatsu:2011, Kim:2011} (see also~\cite{Hwang:2012} for a
critical review of the current cosmological models). In the case of
the CMB the correct extraction of cosmological information crucially
depends on our ability to measure the true CMB 
signal. In practice, the experimentally observed CMB temperature
fluctuations is composed 
of the true CMB signal and foreground signals coming from 
astrophysical sources that emit photons in the frequency ranges
spanned by the observations. The major part of the foreground component
comes from diffuse emissions from our Galaxy, and a small fraction comes
from extra-Galactic point 
sources~\cite{Toffolatti:1997dk} such as radio galaxies and dusty
star-forming galaxies. The Galaxy emissions consist of thermal and
spinning dust emissions, free-free emissions from electrons-ion
scattering,  synchrotron radiation from shock accelerated electrons
interacting with the Galactic magnetic field and a component called
the `haze' whose physical origin is not yet understood. 
These foreground components are usually estimated based on templates
and then subtracted from the observed data~\cite{gold2011}. 

The WMAP data release~\cite{lambdasite} includes masks for our Galaxy
and for extra-Galactic point sources which have been
identified~\cite{gold2011}. Henceforth, we refer to the point sources
mask provided by the WMAP team as PS1. Recently, Scodeller {\em et 
  al.}~\cite{Scodeller:2012fi} reported the detection of new point sources
in the WMAP data that have not been reported before. 
They provide two extended masks~\cite{Scodeller:2012sw}, which we
refer to as PS2 and PS3, and they include
the sources identified by the WMAP team as subsets. PS2 has 1116
sources outside the KQ85 Galactic mask, which were detected either at
5$\sigma$ directly in any of the 5 WMAP channels or at 5$\sigma$ in
internal templates and at 3$\sigma$ in any of the channels. PS3 has
2102 sources outside the KQ85 Galactic mask, which were detected either at
5$\sigma$ directly in any of the 5 WMAP channels or at 5$\sigma$ in
internal templates.   

The goal of this paper is twofold. The first goal is to investigate
whether there is small but statistically significant residual
foreground contamination in the cleaned and masked WMAP
data. Our method is based on calculating
correlations between the foreground field, which has been processed so
as to remove long wavelength
correlations of the galaxy emissions, and the cleaned CMB data. Our
basic premise is that if the there is no residual contamination in the
cleaned and masked data we should obtain no correlation. However, we find
statistically significant positive correlation for WMAP 7 years data
for the Q, V and W differential assemblies (DAs) where we have applied
the KQ75 galactic mask and PS1. We further find that a big fraction
(as big  as 30\% for 
Q channel) comes from regions where the foreground map has large
positive values, which indicates unresolved point sources. 
These results give a clear indication that there are residual
foreground contamination in the cleaned data. A brief report of these 
results  has been presented in~\cite{Chingangbam:2012}. 
We  redo the above calculation of correlation after applying PS2 and
PS3. As is reasonable to expect, we find a decrease in the value of
the correlations and a corresponding decrease in the statistical
significance of  those values, implying that these newly identified
point sources have non-trivial contribution to the correlations.  

Our second goal is to study the effect of the residual
contamination on the estimation of non-Gaussianity parameters by using
Minkowski Functionals (MFs)~\cite{Tomita:1986, Coles:1988, Gott:1990,
  Winitzki:1997jj}. To this end 
we add estimated contaminant fraction to Gaussian CMB simulations and
calculate their effect on the MFs. A comparision between the characteristic
non-Gaussian deviation shapes of the MFS that result from the residual
contamination and the non-Gaussian deviation shapes of WMAP data
using PS1 reveals a remarkable similarity. From this we conclude that the
non-Gaussian deviations seen in MFS measured from WMAP data must come 
predominantly from the residual foreground contamination. 
Further, in order to isolate the effect of the
new point sources contained in PS2 and PS3 we redo the calculation of
non-Gaussian deviation of MFs from WMAP data after masking them. We 
find that the first MF is very strongly affected and the non-Gaussian
deviation shape is completely modified. The effect on the other two
are milder, with the non-Gaussian deviation shapes more or less
unaltered and a decrease in the amplitude of the deviations. This 
can be attributed to the fact that masking the new point sources leads
to a decrease in the level of residual contamination. 
Earlier studies of the effects of contamination on the CMB have mostly
focused  on point sources~\cite{argueso, boughn, babich,
  Lacasa:2011ej}. An investigation of the effect of point sources on
MFs was done in~\cite{Munshi:2012kq}.  

This paper is organized as follows. In section 2 we present
calculations of the correlations between the foreground and
cleaned CMB maps and their statistical significance after applying
point sources masks PS1, PS2 and PS3. In section 3, we compute
MFs from Gaussian simulations to which a fraction of the foreground
field is added and compare the non-Gaussian deviations to the
corresponding deviations measured from WMAP 7 years data using PS1. 
We further study the effect of masking the additional
point sources in PS2 and PS3 on the MFs. We end with concluding
remarks in  section 4. 

\section{Quantifying residual foreground contamination}

We begin with the expectation that any two random fluctuation fields
that originate from completely different physical processes will not
have any correlation. 
Let $f$ and $f'$ be two random fields that have zero mean
values, defined  on the  surface of a two dimensional sphere. Let their rms
values be denoted by $\sigma_0$ and $\sigma_0'$,
respectively. By rescaling them as  
$\nu(i) \equiv f(i)/\sigma_0$ and $\nu'(i) \equiv f'(i)/\sigma_0'$, 
where $i$ denotes the pixel number, we can define a correlation
parameter, $r$, as  
\be
r \equiv <\nu(i)\,\nu'(i)>,
\ee
where the bracket denotes average over all pixels. We expect $r$ to be
zero if the two fields are uncorrelated and non-zero otherwise. In
numerical calculations we will always get a non-zero value
of $r$ even for two fields that are known to be uncorrelated, due to
the finite number of pixels, and we need to further test statistically
whether the value is small enough to be be considered as practically
zero.

The observed WMAP data, $f^{\rm obs}$, is a sum of the true CMB signal
and foreground contamination. Let us call the foreground component
that is estimated using a combination of galaxy observation and
theoretical modeling, as the `apparent' foreground field, denoted by
$f^{\rm appfg}$, keeping in mind that there may be small error in its
estimation. This field is then subtracted pixel by pixel from $f^{\rm
  obs}$ to leave behind the `cleaned' CMB signal, which we denote by
$f^{\rm cleaned}$. By definition, $f^{\rm cleaned}$ has zero mean. If
$f^{\rm appfg}$ has been correctly estimated, then we expect it to
have negligibly small correlation with $f^{\rm cleaned}$, since they
come from totally different physical processes. However, if the
estimation is on the right track but not fully correct, then we should
expect some residual contamination in the signal field. This should
show up as non-zero correlation between $f^{\rm cleaned}$ and $f^{\rm
  appfg}$. 

\subsection{Peak field}

\begin{figure}[h]
\begin{center}
\resizebox{2.8in}{1.3in}{\includegraphics{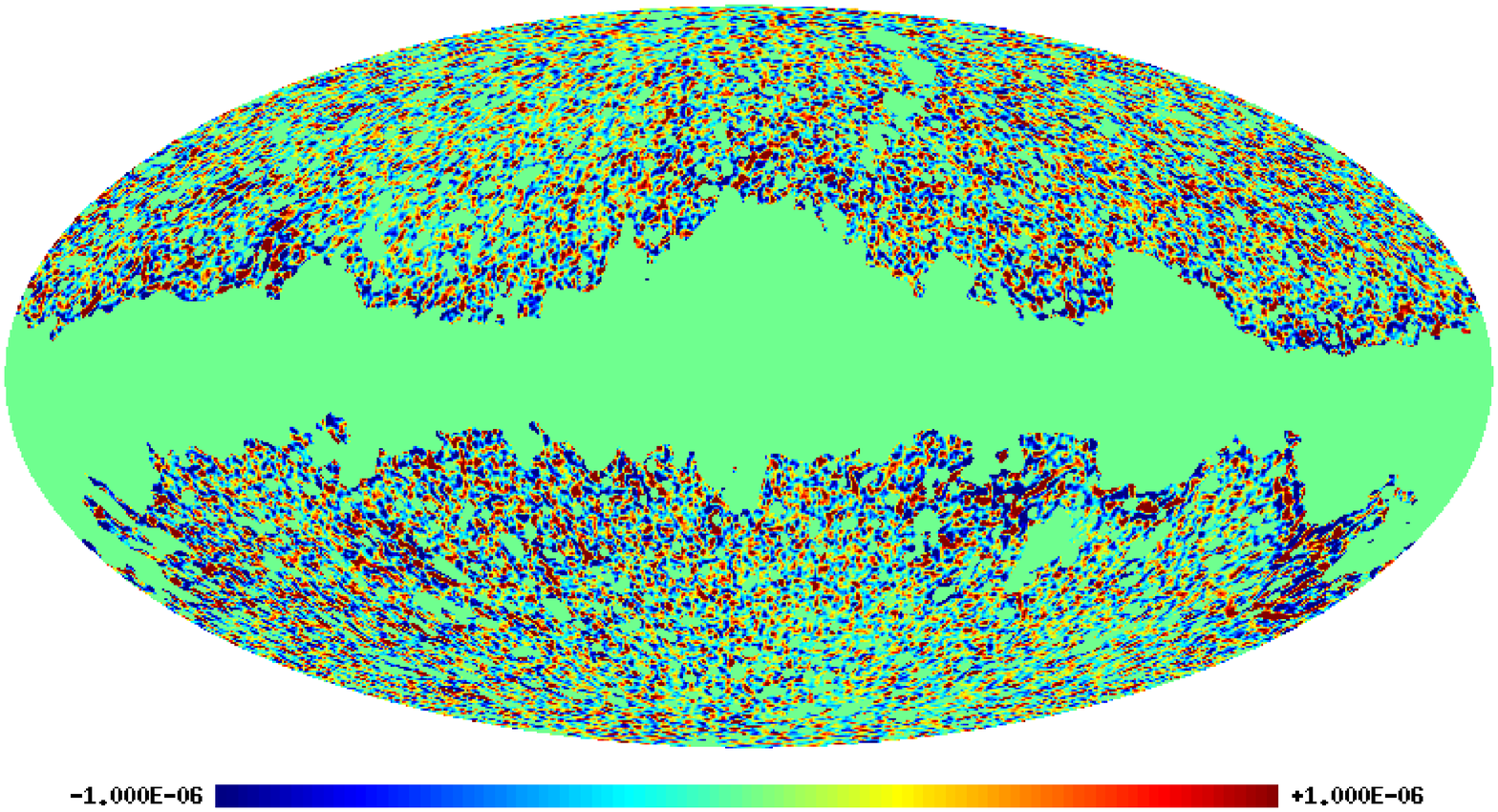}}
\hskip .5cm
\resizebox{2.8in}{1.3in}{\includegraphics{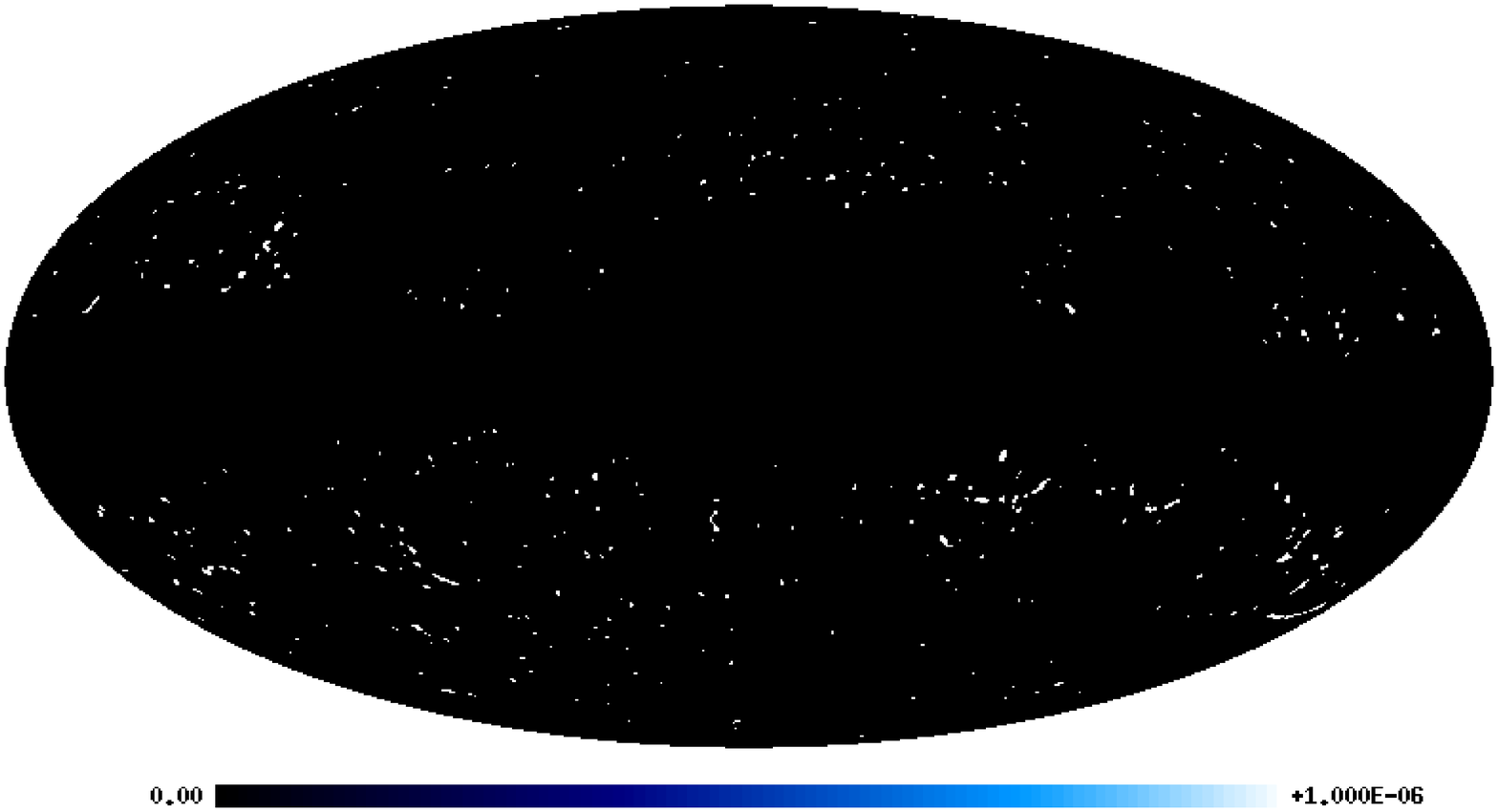}}
\end{center}
\caption{{\em Left panel}: Peak field for ${\rm Q}_1$ DA. {\em Right
    panel}: Locations of pixels where the same peak field has values
  above   $3\sigma^{\rm peak}$ are shown in white. }
\label{fig:peak_Q1}
\end{figure}

Our analysis is done using the 7 years data from the eight
differential assemblies (DAs) of WMAP, namely, ${\rm Q}_1$, ${\rm
  Q}_2$, ${\rm V}_1$, ${\rm V}_2$, ${\rm W}_1$, ${\rm W}_2$, ${\rm
  W}_3$ and ${\rm W}_4$.  
For each DA the Galactic foreground is obtained from  
\begin{equation}
f^{\rm appfg} = f^{\rm obs} - f^{\rm cleaned},
\label{eqn:foreground}
\end{equation}
where `appfg' indicates that $f^{\rm appfg}$ is the `apparent'
foreground and a fraction of it may be left behind in $f^{\rm cleaned}$.  
To examine the existence of the residual foreground in the cleaned map
we will compare the cleaned map with the foreground maps where the
large-scale variation of the Galactic emissions is removed. A
foreground map with the large-scale variation subtracted is called the
{\em peak field} and is defined as
\begin{equation}
f^{\rm peak} \equiv \left( f^{{\rm appfg},\theta_s} -  f^{{\rm
    appfg},3\theta_s}  \right)  
- <f^{{\rm appfg},\theta_s} -  f^{{\rm appfg},3\theta_s}>,
\end{equation}
where $\theta_s$ and $3\theta_s$ are FWHM values at which we have
smoothed the field. By definition $f^{\rm peak}$ has zero mean. The
left panel of Fig.~(\ref{fig:peak_Q1})  shows the peak field for Q1
channel. In the right panel we have shown pixels (in white) of the same peak
field which have values above $3\sigma^{\rm peak}$. 
 
To reduce the boundary effects by controlling the degree of masking we
use the foreground mask map smoothed over FWHM$=3\theta_s$. The pixels
of this map have value one well inside the mask boundaries and zero
well outside, but have values between zero and one near the
boundaries. The distance from the original mask boundaries is then
encoded in the pixel values of the smoothed mask map. By using
some threshold value, $s_{\rm mask}$, of the smoothed mask pixels, we
can control how far away we stay away from the mask boundary. Staying
$2\sigma$ away from the boundary corresponds to choosing pixels with
$s_{\rm mask}> 0.89$. As we choose larger values of $s_{\rm mask}$ we
stay further away from the boundaries and the sky fraction decreases. 

\subsection{Correlation between peak and cleaned CMB fields}

Let us denote
\be
\nu^{\rm cleaned}(i) \equiv \frac{f^{\rm cleaned}(i)}{\sigma^{\rm
    cleaned}}, \quad \nu^{\rm peak}(i) \equiv \frac{f^{\rm
    peak}(i)}{\sigma^{\rm peak}}, 
\ee
and  define
\be
r_c \equiv <\nu^{\rm cleaned}\,\nu^{\rm peak}>_{\theta_s}, 
\ee
where the suffix $\theta_s$ is to remind us that we do the calculation
for a choice of FWHM at which $f^{\rm cleaned}$ has also been smoothed. 
$\sigma^{\rm cleaned}$ and $\sigma^{\rm peak}$ are of the orders of
$10^{-5}$ and $10^{-7}$, respectively.  
\renewcommand{\arraystretch}{1.4}
\begin{table}
\begin{center}
\begin{tabular}{|p{2.3cm}|c|c|c|c|c|c|c|c|}
\cline{2-9}
\multicolumn{1}{p{2.3cm}|}{ {}}
& $Q_1$ & $Q_2$ & $V_1$  & $V_2$ & $W_1$  & $W_2$ & $W_3$  & $W_4$ \\ 
\hline
\multirow{2}{3cm} {$r_c$ for $s_{\rm mask}=0.89$} & $0.026$ &  $0.025$  &  $0.020 $  & $0.019 $ &  $0.010$  &  $0.008 $ & $0.009 $ &  $0.006$ \\ \cline{2-9}
& $0.018 $ &  $0.018$  &  $0.017$  & $0.016$ &  $0.009$  &  $0.008$ & $0.007$ &  $0.007$ \\ 
\hline
\multirow{2}{3cm} {$r_c$ for $s_{\rm mask}=0.91$} & $0.025 $ &
$0.025$  &  $0.020 $  & $0.019$  & $0.009$  &  $0.007$ & $0.008$ &  $0.006$ \\  \cline{2-9}
& $0.018 $ &  $0.017$  &  $0.016 $ & $0.016 $ &  $0.008 $  &  $0.008 $ & $0.006$ &  $0.006$ \\  \cline{2-9}
\hline
\multirow{2}{3cm} {$r_c$ for $s_{\rm mask}=0.93$} & $0.025$ &  $0.024$
&  $0.019$  & $0.018 $ &  $0.008$  &  $0.007$ & $0.007$ &  $0.004$ \\  \cline{2-9}
& $0.017$ &  $0.017$  &  $0.016 $ & $0.015 $&  $0.007$  &  $0.007$ & $0.005$ &  $0.006$ \\ \cline{2-9}
\hline

\end{tabular}
\caption{\small{$r_c$ values for the eight DAs are
  shown. Point sources mask used is PS1. For each DA and $s_{\rm
    mask}$, the upper value gives $r_c$ calculated   using all
  unmasked pixels, while the lower value has been calculated after 
  excluding pixels having $\nu^{\rm peak} > 3$ also. The sky fractions
  for the three  $s_{\rm mask}$ values from top to bottom are roughly
  $62\%$, $60\%$ and  $58\%$. }} 
\label{table:rctable}
\end{center}
\end{table}
Table (\ref{table:rctable}) summarizes the main results for $r_c$
where point sources mask PS1 has been used. We have chosen
$\theta_s=35'$ based on the resolution of Q1 channel. Two values of
$r_c$ are shown for each  
DA and $s_{\rm mask}$. The upper value is the case where $r_c$ is
calculated using all unmasked pixels, while the  
lower value is the case where pixels with $\nu^{\rm peak} > 3$, shown
in the right panel of Fig.~(\ref{fig:peak_Q1}) for Q1, have been
excluded. The first observation we make is that $all$ $r_c$ values are
positive. For Q channels {\em we get considerably larger correlation
  when we keep all unmasked pixels, larger by about 30\%. This
  indicates that there is non-trivial correlation arising from the
  pixels  with $\nu^{\rm peak} > 3$.} For V channels the difference is
about 20\% while W channels don't seem to be affected. The sky
fractions for the three  $s_{\rm mask}$ values are roughly, $62\%$,
$60\%$ and  $58\%$, respectively. For Q and V channels, as we stay
further away from the mask boundaries there is small but systematic
decrease of $r_c$.  

\subsection{Statistical significance of $r_c$ values}
\label{sec:statsig}
\renewcommand{\arraystretch}{1.2}
\begin{center}
\begin{table}
\begin{minipage}[b]{0.7\linewidth}\centering
\begin{tabular}{|p{1.cm}|r|r|r|r|r|r|r|r|}
\cline{2-9}
\multicolumn{1}{p{2.2cm}|}{ {}} & $Q_1$  & $Q_2$ & $V_1$  & $V_2$ &
$W_1$  & $W_2$ & $W_3$  & $W_4$ \\ 
\hline
\multirow{2}{2.2cm} {$N$ for $s_{\rm mask}=0.89$}   & 0  & 0  &  5 &
8& 120 & 169 & 139 & 224 \\ \cline{2-9}
& 14  & 14  &  105 & 114 & 276 & 276 & 314 & 303 \\ \cline{2-9}
\hline
\multirow{2}{2.2cm} {$N$ for $s_{\rm mask}=0.91$}   & 0  & 0  &  7 &
10 & 136 & 193 & 160 & 255\\ \cline{2-9}
& 15  & 16  &  110 & 122 & 284 & 286 & 338 & 318 \\ \cline{2-9}
\hline
\multirow{2}{2.2cm} {$N$ for $s_{\rm mask}=0.93$}   & 0  & 0  &  14 &
14& 161 & 219 & 199 & 294 \\ \cline{2-9}
& 18  & 20  &  121 & 130 & 295 & 294 & 350 & 342 \\ \cline{2-9}
\hline
\end{tabular}
\end{minipage}
\hspace{.4cm}
\begin{minipage}[b]{0.2\linewidth}
\begin{tabular}{|p{2.2cm}|r|}
\hline
\multirow{2}{3cm} {$N_0$ for $
s_{\rm mask}=0.89$}   & 0\\ \cline{2-2}
& 13\\
\hline
\multirow{2}{3cm} {$N_0$ for $
s_{\rm mask}=0.91$}   & 0\\ \cline{2-2}
& 15\\
\hline
\multirow{2}{3cm} {$N_0$ for $
s_{\rm mask}=0.93$}   & 0\\ \cline{2-2}
& 18\\
\hline
\end{tabular}
\end{minipage}
\caption{\small{{\em Left:} Number of maps, $N$, having $r_g>r_c$ for
    individual DAs, out of 1000 Gaussian maps for PS1. As in
    Table~(\ref{table:rctable}), upper values are for all unmasked
    pixels included, while lower values are for the case when pixels
    with $\nu^{peak} > 3$ have also been excluded. {\em Right:} Number
    of Gaussian maps, $N_0$, having $r_g>r_c$ simultaneously for all
    DAs.}} 
\label{table:Ntable}
\end{table}
\end{center}

We investigate how likely it is to get the observed $r_c$ values given
in Table (\ref{table:rctable})  by comparing with correlations between
the peak field and 
Gaussian CMB simulations. For this purpose we simulate 1000 Gaussian
CMB maps with WMAP 7 years parameter values, add pixel window effect,
beam smearing and WMAP 7 years noise characteristics. Next we smooth
by FWHM 35' and mask in exactly the same way as we did when
calculating $r_c$ and calculate the correlation with the peak
field. We denote the correlation value by $r_g$. The Gaussian fields
are uncorrelated with the signal field and we should get small value
of $r_g$. This exercise will tell us what is the typical value of
`small' $r_c$ that we can approximate to be zero for the number of
pixels under consideration and how likely are our observed  $r_c$
values to occur by random fluctuation and not due to a true
correlation.   

We count, out of the thousand $r_g$ values, how many are
greater than $r_c$. The results are shown in
Table~(\ref{table:Ntable}). The left table shows the number, $N$, of
Gaussian maps having $r_g > r_c$ for each individual DA, for the three
$s_{\rm mask}$ values used earlier for calculating $r_c$ and
including/excluding the pixels having $\nu^{\rm peak} > 3$.  
When all unmasked pixels are included, we get $N=0$ for all $s_{\rm
  mask}$ values for Q channel, for V channels $N$ lies between 5 and
14, while for $W$ channels $N$ lies between 120 and 294. These numbers
imply that the $r_c$ values for Q and V are statistically significant,
whereas, the values for W channels have much lower significance. When
pixels with $\nu^{\rm peak} > 3$ are also excluded, we find a
reduction of $N$ for all the channels. 
The table on the right side of Table~(\ref{table:Ntable}) shows the
number, $N_0$, of Gaussian maps having $r_g > r_c$ {\em simultaneously
  for all DAs}. These values  are again significant. Therefore, we
conclude that the cleaned WMAP data, particularly Q and V channels,
contain small but statistically significant amount of residual
foreground contamination.  

\vskip .5cm
\subsection{Correlation between peak and cleaned CMB fields after 
  applying extended point sources masks PS2 and PS3}

\renewcommand{\arraystretch}{1.4}
\begin{table}
\begin{center}
\begin{tabular}{|p{2.3cm}|c|c|c|c|c|c|c|c|}
\cline{2-9}
\multicolumn{1}{p{2.3cm}|}{ {}}
& $Q_1$ & $Q_2$ & $V_1$  & $V_2$ & $W_1$  & $W_2$ & $W_3$  & $W_4$ \\ 
\hline
\multirow{2}{3cm} {$r_c$ for $s_{\rm mask}=0.89$} & $0.013$ &  $0.013$  &  $0.010 $  & $0.010 $ &  $0.002$  &  $0.001 $ & $0.001 $ &  $-0.001$ \\ \cline{2-9}
& $0.010 $ &  $0.009$  &  $0.009 $  & $0.009$ &  $0.005$  &  $0.005$ & $0.003$ &  $0.003$ \\ 
\hline
\multirow{2}{3cm} {$r_c$ for $s_{\rm mask}=0.91$} & $0.013 $ &
$0.013$  &  $0.010 $  & $0.009$  & $0.002$  &  $0.001$ & $0.001$ &  $-0.002$ \\  \cline{2-9}
& $0.010 $ &  $0.009$  &  $0.009 $ & $0.009 $ &  $0.005 $  &  $0.004 $ & $0.002$ &  $0.002$ \\  \cline{2-9}
\hline
\multirow{2}{3cm} {$r_c$ for $s_{\rm mask}=0.93$} & $0.012$ &  $0.012$
&  $0.009$  & $0.009 $ &  $0.001$  &  $-1\times 10^{-6}$ & $-4\times 10^{-5}$ &  $-0.003$ \\  \cline{2-9}
& $0.009$ &  $0.009$  &  $0.009 $ & $0.008 $&  $0.004$  &  $0.003$ & $0.001$ &  $0.002$ \\ \cline{2-9}
\hline

\end{tabular}
\caption{\small{$r_c$ values for the eight DAs after applying
    PS2. Y As in Table~(\ref{table:rctable}), for each DA and $s_{\rm
      mask}$, the upper value gives $r_c$ 
    calculated   using all  unmasked pixels, while the lower value has
    been calculated after excluding pixels having $\nu^{\rm peak} > 3$
    also. The sky fractions for the three  $s_{\rm mask}$ values from
    top to bottom are roughly $61\%$, $59\%$ and  $57\%$. }}
\label{table:rctable_hansen}
\end{center}
\end{table}
\renewcommand{\arraystretch}{1.2}
\begin{center}
\begin{table}

\begin{minipage}[b]{0.7\linewidth}\centering
\begin{tabular}{|p{1.cm}|r|r|r|r|r|r|r|r|}
\cline{2-9}
\multicolumn{1}{p{2.2cm}|}{ {}} & $Q_1$  & $Q_2$ & $V_1$  & $V_2$ &
$W_1$  & $W_2$ & $W_3$  & $W_4$ \\ 
\hline
\multirow{2}{2.2cm} {$N$ for $s_{\rm mask}=0.89$}   & 37  & 39  &  148 &
159 & 407 & 462 & 448 & 549 \\ \cline{2-9}
& 157  & 163  &  269 & 280 & 384 & 396 & 442 & 430 \\ \cline{2-9}
\hline
\multirow{2}{2.2cm} {$N$ for $s_{\rm mask}=0.91$}   & 39  & 43  &  158 &
167 & 430 & 482 & 475 & 584\\ \cline{2-9}
& 164  & 167  &  285 & 292 & 392 & 407 & 458 & 448 \\ \cline{2-9}
\hline
\multirow{2}{2.2cm} {$N$ for $s_{\rm mask}=0.93$}   & 45  & 46  &  166 &
168 & 449 & 505 & 511 & 628 \\ \cline{2-9}
& 173  & 175  &  293 & 301 & 406 & 416 & 477 & 470 \\ \cline{2-9}
\hline
\end{tabular}
\end{minipage}
\hspace{.4cm}
\begin{minipage}[b]{0.3\linewidth}
\begin{tabular}{|p{2.2cm}|r|}
\hline
\multirow{2}{3cm} {$N_0$ for $
s_{\rm mask}=0.89$}   & 35\\ \cline{2-2}
& 148\\
\hline
\multirow{2}{3cm} {$N_0$ for $
s_{\rm mask}=0.91$}   & 38\\ \cline{2-2}
& 154\\
\hline
\multirow{2}{3cm} {$N_0$ for $
s_{\rm mask}=0.93$}   & 43\\ \cline{2-2}
& 162\\
\hline
\end{tabular}
\end{minipage}
\caption{\small{{\em Left:} Number of maps, $N$, having $r_g>r_c$ for
    individual DAs, out of 1000 Gaussian maps, calculated after
    applying PS2. As in Table~(\ref{table:Ntable}), upper values are
    for all unmasked pixels included, while lower values are for the
    case when pixels with $\nu^{peak} > 3$ have also been
    excluded. {\em Right:} Number of Gaussian maps, $N_0$, having
    $r_g>r_c$ simultaneously for all DAs.}} 
\label{table:Ntable_hansen}
\end{table}
\end{center}
We have repeated the calculation of $r_c$ and the analysis of the
statistical significance after masking after applying PS2. Masking the
new point sources results in a further decrease of roughly $1\%$ of
the sky fraction. The $r_c$ values that we obtain are shown in Table
(\ref{table:rctable_hansen}). The significance test results are shown
in Table (\ref{table:Ntable_hansen}). We find a clear decrease in the
correlation values and their statistical significance which indicates
that there is reduction in the residual contamination, as should be
expected .  

\vskip .5cm
\section{Minkowski Functionals and residual foreground}

The morphological properties of excursions sets of the CMB (the set of
all pixels having temperature fluctuation values greater than or equal
to some threshold value, $\nu$) can be neatly captured by the so called
Minkowski Functionals. There are three MFs that are relevant for the
CMB. The first is the area fraction, $V_0$, of the excursion set, the
second is the total length, $V_1$, of iso-temperature contours or
boundaries of the excursion sets and the third  is the genus, $V_2$,
which is the difference between the numbers of hot and cold
spots~\cite{Gott:1990}.    
For a Gaussian random field the MFs are given by, 
\begin{equation}
\label{eq:mf}
V_k(\nu)=A_k \, H_{k-1}(\nu)\,e^{-\nu^2/2}, \quad k=0,1,2. 
\end{equation}
$H_n(\nu)$ is the $n$-th Hermite polynomial and the amplitude $A_k$
depends only on the angular power spectrum 
$C_l$. It is given by 
\begin{equation}
\label{eq:ak}
A_k=\frac1{(2\pi)^{(k+1)/2}}\frac{\omega_2}{\omega_{2-k}\omega_k}
\left(\frac{\sigma_1}{\sqrt{2}\sigma_0}\right)^k,
\end{equation}
\begin{equation}
\label{eq:vk}
\sigma_j^2\equiv \frac1{4\pi}\sum_l(2l+1)\left[l(l+1)\right]^j C_l
W^2_l, 
\end{equation}
with $\omega_k\equiv \pi^{k/2}/{\Gamma(k/2+1)}$.  
$\sigma_1$ is the rms of the gradient of the field and $W_l$
represents the smoothing kernel determined by the pixel and beam
window functions and any additional smoothing. The presence of any
small deviation from Gaussianity will appear as deviations from these
formulas. The MFs are useful because they have characteristic 
non-Gaussian deviation shapes for different types of non-Gaussianity
and can distinguish them. They carry information of all orders of
$n$-point functions and this makes them unbiased towards specific forms
of non-gaussianity. 

For the numerical computation of MFs for any given random field we use the
method described in~\cite{Schmalzing:1997uc}. This method was shown to
have numerical inaccuracies which are of specific forms arising from
the finite approximation of the delta function and which scales 
as the square of the finite binning of the temperature threshold
values at leading order~\cite{Lim:2012}. In our calculations we
estimate and subtract  these inaccuracies and we denote the corrected
result by $V_i^{NG}$.  For weakly non-Gaussian fields we can obtain
the Gaussian component by using the formula Eq.~(\ref{eq:mf}) where
the amplitude is computed by measuring $\sigma_0$ and $\sigma_1$
directly from the field. We denote it by $V_i^{G}$. The non-Gaussian
deviation is then given by 
\begin{equation}
\Delta V_i \equiv V_i^{NG} - V_i^{G}. 
\label{eqn:dmf}
\end{equation}

\subsection{Effect of residual contamination on Minkowski Functionals} 

\begin{figure}[h]
\begin{center}
\resizebox{2.0in}{2.in}{\includegraphics{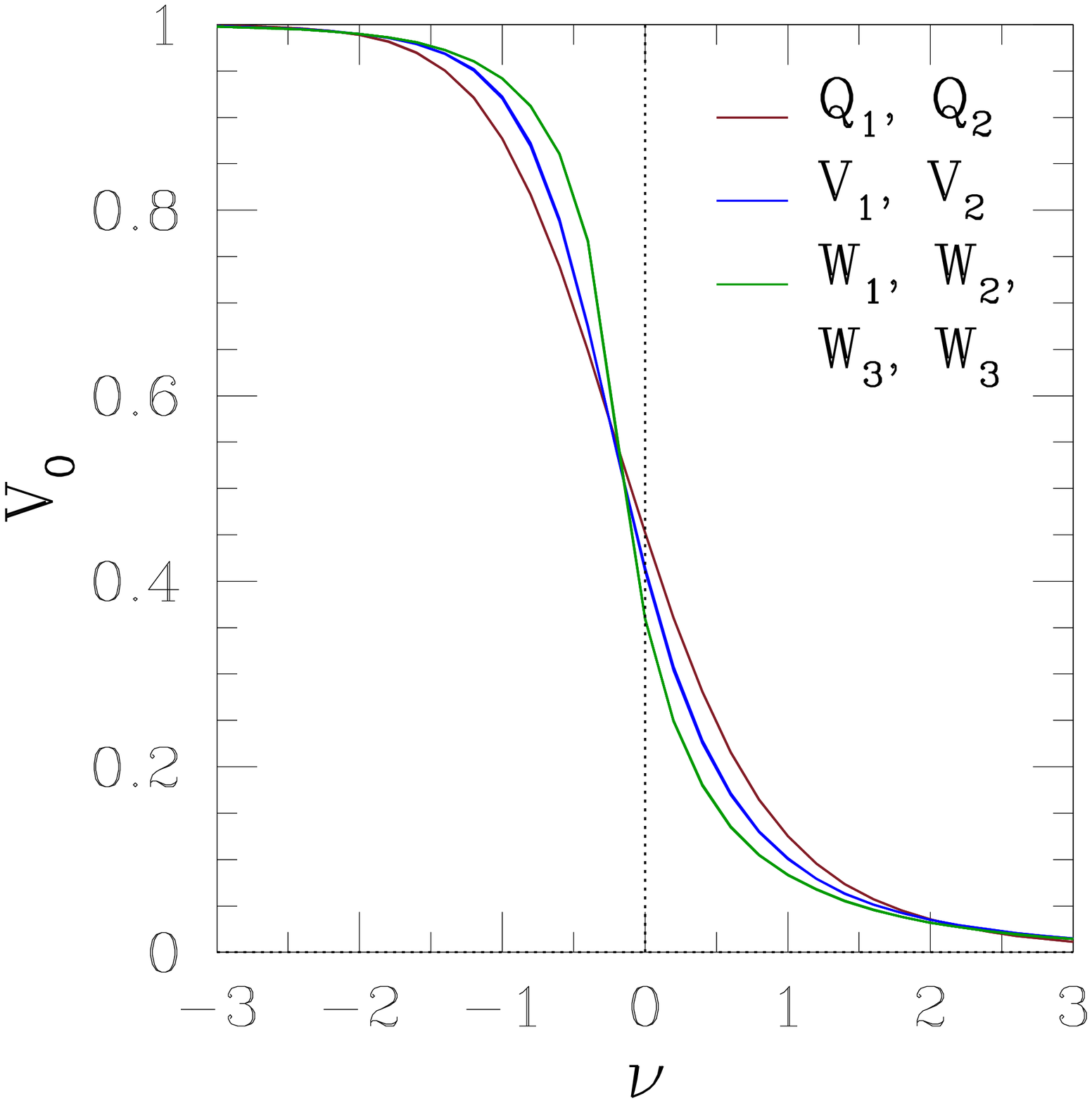}}
\resizebox{2.0in}{2.in}{\includegraphics{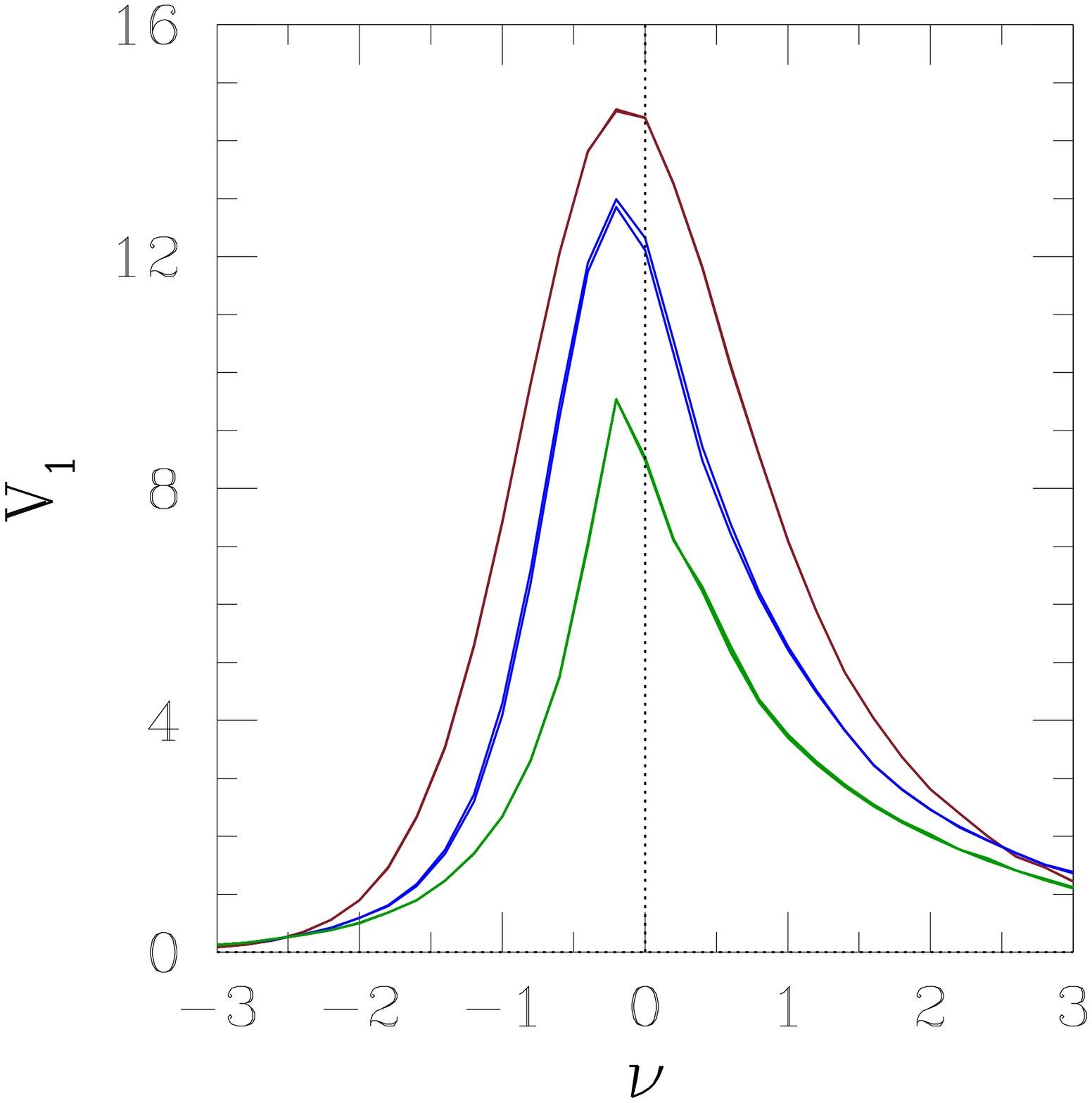}}
\resizebox{2.0in}{2.in}{\includegraphics{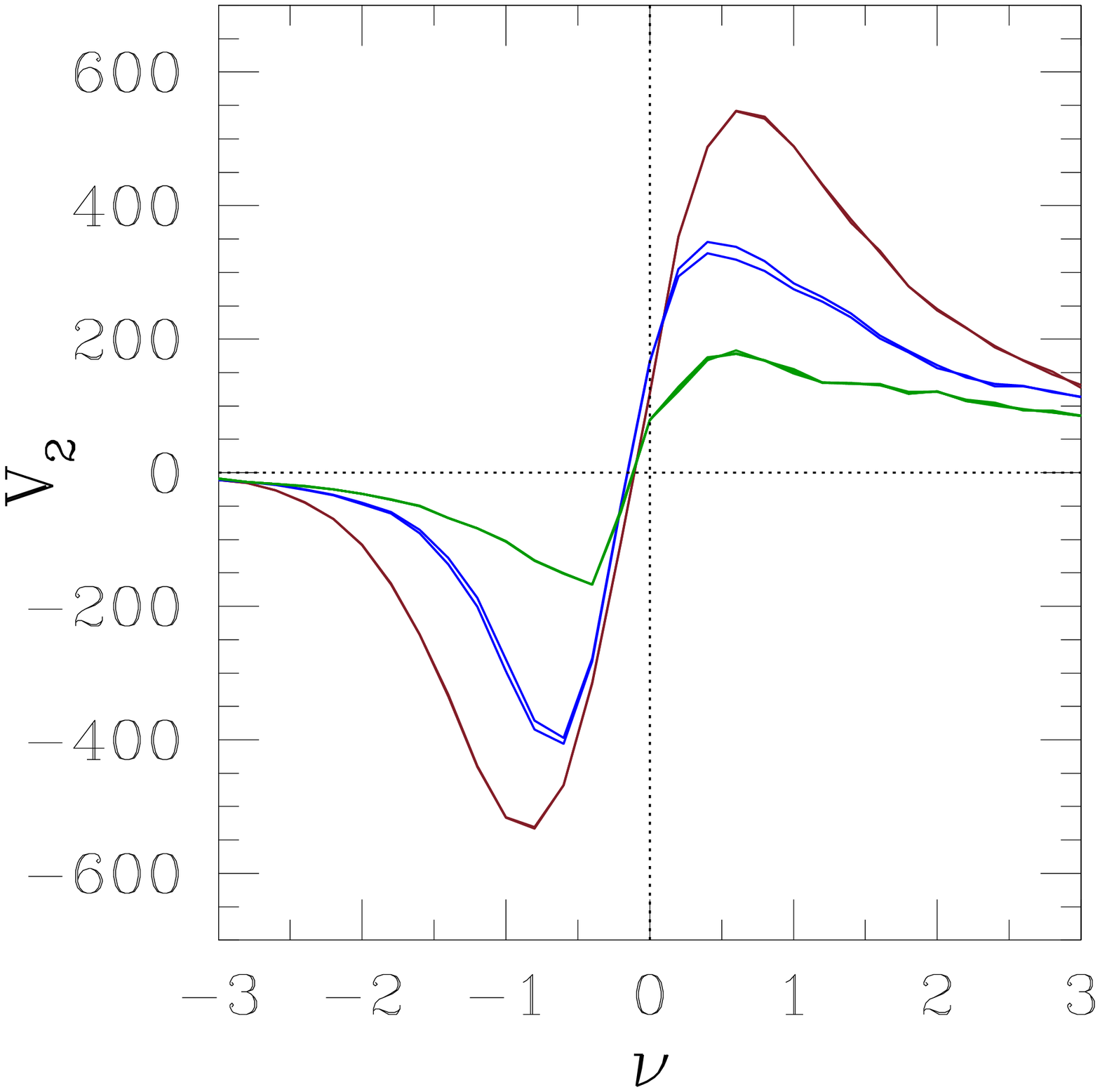}}
\end{center}
\caption{Minkowski functionals for the peak fields.}
\label{fig:MF_peak}
\end{figure}
In this subsection we study how the residual foreground  contamination
affects the MFs. We begin by examining the shapes of the MFs for the
peak fields shown in Fig.~(\ref{fig:MF_peak}) for
each of the DAs. It is obvious that they have strong departures from 
Gaussian shapes. We can therefore expect that if any small fraction of
the peak fields contaminate the CMB field it will show up as
non-Gaussian deviation in the MFs. 

\begin{figure}[h]
\begin{center}
\resizebox{5.in}{6in}{\includegraphics{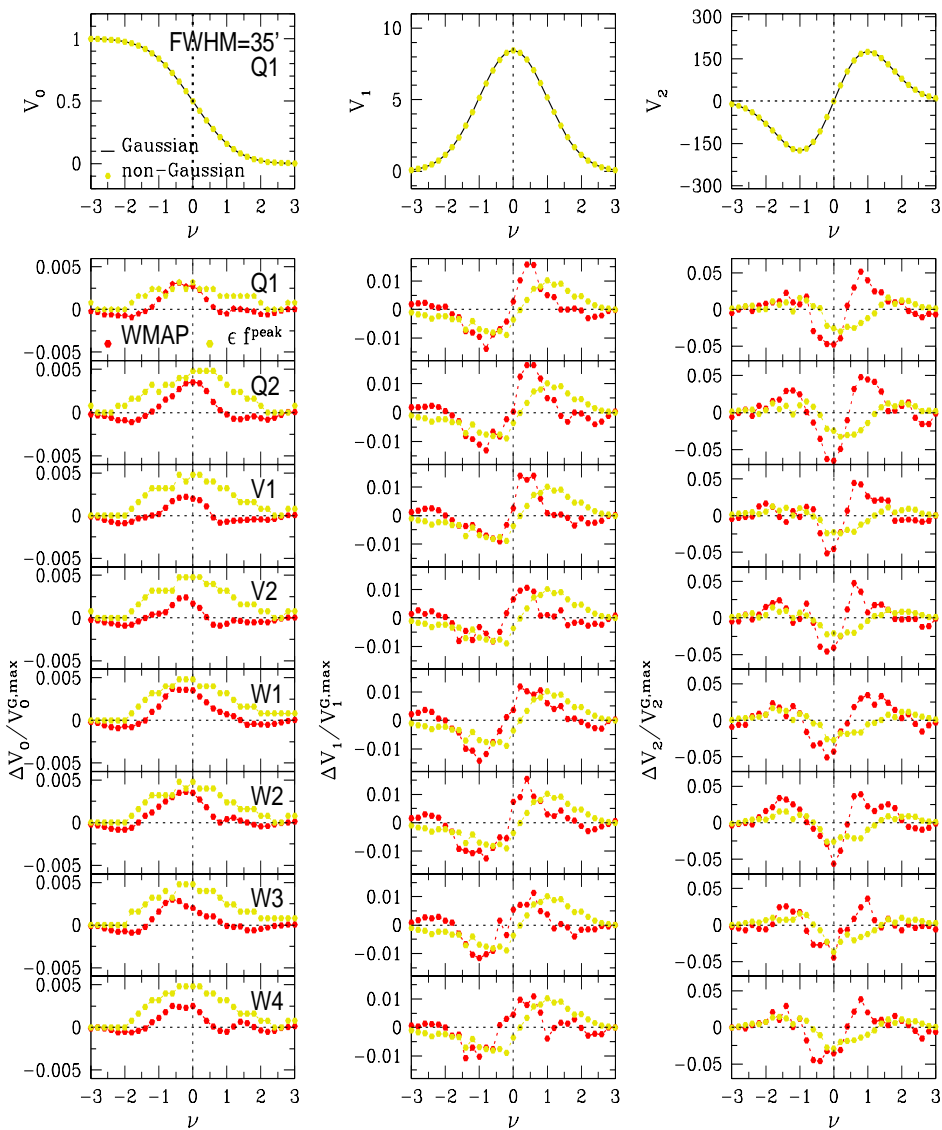}}
\end{center}
\caption{Minkowski Functionals and their non-Gaussian deviations
  (Eq.~(\ref{eqn:dmf})), measured  
  from Gaussian simulations to which residual contaminant fraction has
  been added, given by Eq.~(\ref{eqn:fcontaminated}). Average over
  1000 simulations.}  
\label{fig:MF_contaminated}
\end{figure}
In order to mimic and quantify the effect of the residual contamination
on the MFs we add $\epsilon f^{\rm peak}$ to Gaussian simulated maps, as,
\begin{equation}
f^{\rm contaminated} = f^G + \epsilon f^{\rm peak},
\label{eqn:fcontaminated}
\end{equation}
where $f^{\rm G}$ is the simulated Gaussian map, to which we have added
instrumental effects, as descrived in section
(\ref{sec:statsig}). Note that the 
largest contribution to the non-Gaussian deviation of the MFs arising
from non-zero $\epsilon$ will scale linearly with
it~\cite{Hikage:2006fe}. The MFs and their non-Gaussian deviations
(yellow dots) computed from $f^{\rm contaminated}$,  
averaged over 1000 maps, are shown in Fig.~(\ref{fig:MF_contaminated}).  
We have chosen values of $\epsilon$ which result in amplitudes of the MFs
similar to the observed ones. The $\epsilon$ value used for these plots is
$\epsilon=8r_c/(r_c+\sigma^{\rm peak} /\sigma^{\rm  cleaned})$ for each 
respective DA. In the same figure we have also shown the $\Delta V_i$
computed from the WMAP 7 years data (red dots) after applying PS1. As
seen in the figure, there is remarkable agreement between the two
plots.  We infer that most of the non-Gaussian
deviation  that we measure in the WMAP data is contributed by residual
foreground contamination. 

\subsection{Effect of PS2 and PS3 on Minkowski Functionals from WMAP}

\begin{figure}
\begin{center}
\resizebox{5.in}{6.in}{\includegraphics{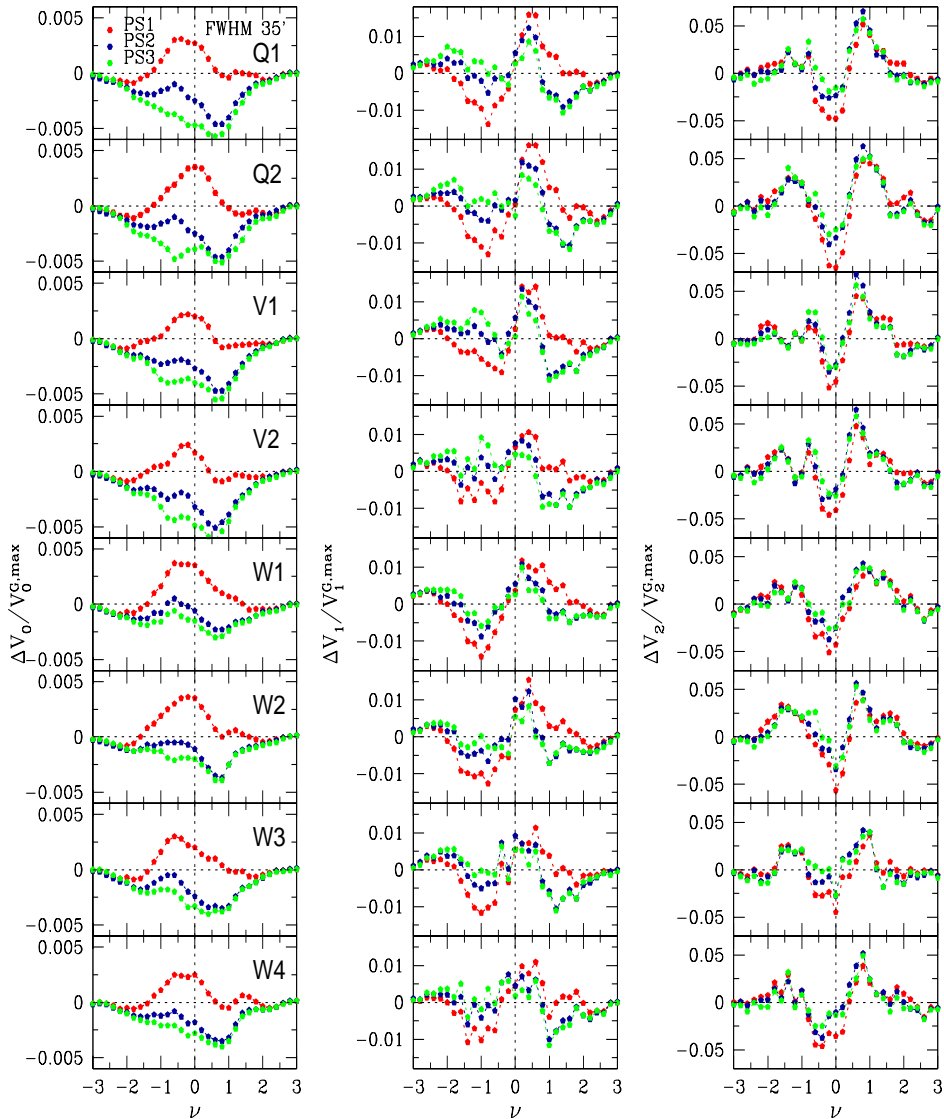}}
\end{center}
\caption{Non-Gaussian deviations of Minkowski functionals for
  the eight DAs of WMAP 7 years data for the three 
  cases where point sources masks PS1, PS2 and PS3 were applied. }
\label{fig:MF_wmap7}
\end{figure}

We calculate the MFs for WMAP 7 years data after applying PS2 and PS3. 
The non-Gaussian deviations are shown in~Fig.(\ref{fig:MF_wmap7}), along
with the result of PS1 so as to compare the three.  The Galaxy mask
applied is KQ75 as done in previous sections. We find that $\Delta
V_0(\nu)$ is strongly affected by the removal of the new 
point sources, it flips sign. This is simple to understand, as
explained below. Let $N$ denote the total number of unmasked
pixels. At any $\nu$ let us denote the number
of pixels greater than or equal to $\nu$ by $n(\nu)$.  $V_0(\nu)$ is
given by  $n(\nu)/N$. When we mask new point sources we exclude, say
$m$, positive valued pixels. Then the effect of the new masking gives
new value $V_0'(\nu)$. Since $m$ is positive, this implies $V_0'(\nu)
< V_0(\nu)$.  If we started with $V_0$ which is greater than the
Gaussian  value and this decrease makes $V_0'$ less than the
Gaussian value, then $\Delta V_0$ will flip sign which is the case
here. This suggests that $V_0(\nu)$ is unreliable for
extraction of non-Gaussianity information due to our imprecise
knowledge of point sources in the sky. 
For $\Delta V_1(\nu)$ the amplitude is decreased considerably but the
shape is more or less unaffected. The genus is affected the least and
the main effect is a reduction of the non-Gaussian deviation around
$\nu=0$. These effects are due to the reduction in the level of
contamination due to the masking of the new point sources. 

\section{Conclusion}

We have analysed the cleaned WMAP 7 years data with the goal of
quantifying the amount of residual foreground contamination outside
the Galactic and point sources masks and the
resulting bias in the estimates of primordial non-Gaussianity by using 
Minkowski Functionals. The presence of significant residual
contamination is confirmed by calculations of correlations between the
cleaned maps and the foreground maps which give values that are found
to be statistically significant. The Q channel is found to have the
strongest correlations and hence largest residual contamination while
W channel has the least. For Q and V channels we found that a big
fraction of the contamination come from  pixels where the foreground
fields have large values. A comparision of the correlation and
significance values obtained after applying the point sources mask
provided by WMAP and the extended masks of Scodeller {\em et
  al}. reveals that the extended masks remove some fraction of the
residual contamination, as should be expected.  

The above results have important implications for the extraction of
cosmological parameters from observational data, particularly on
the search for primordial non-Gaussianity, using  the cleaned WMAP
data. In order 
to  understand the implications we simulate contaminated CMB maps by
adding a fraction of the foreground field to Gaussian maps and measure
Minkowski Functionals from them. The non-Gaussian
deviation shapes of all the three MFs are found to have remarkable
agreement with what is measured from the cleaned WMAP data after
applying PS1. 
Non-Gaussian deviations of MFs calculated after applying PS2 and PS3
give a reduction in the overall magnitude of the deviations compared
to PS1. $\Delta V_0$ actually changes sign owing to
its strong sensitivity to point sources and is not reliable to be used
for constraining primordial non-Gaussianity. The shapes of $\Delta
V_1$ and $\Delta V_2$ are relatively insensitive to the different
masking with the main effect being a decrease in the amplitude. These
results are consistent with a reduction in the level of residual
contamination. 

We conclude that the cleaned WMAP 7 years data contains significant
amount of residual foreground contamination,  both from diffuse
Galactic emissions and unresolved extra-Galactic point sources. Note
that more than 15000 point sources have already been
identified from data from the PLANCK satellite~\cite{PLANCK}. A 
rough visual comparision between the amplitudes of $\Delta V_i$ in
Figs.~(\ref{fig:MF_contaminated}) and (\ref{fig:MF_wmap7}) and the
corresponding amplitudes of deviations due to the local primordial
non-Gaussianity parameter $f_{\rm NL}$~\cite{Hikage:2006fe} tells us
that the cleaned data contains residual contamination of similar
levels as $f_{\rm NL} > 100$. Unless this is removed the constraints put
on primordial non-Gaussianity parameters, such as
in~\cite{Hikage:2012bs}, are not sensible. Our 
calculations may be refined to obtain a good estimate of the residual
contamination fraction encoded in the parameter $\epsilon$. It can
then be further subtracted from the cleaned data and the resulting
maps can be used to search for primordial non-Gaussianity. 
It is also imperative that we redo such analysis after
applying the extended point sources masks PS2 and PS3. We are
currently initiating this investigation.  

\section*{Acknowledgment}
We thank Korea Institute for Advanced Study for providing computing
resources (KIAS Center for Advanced Computation Linux Cluster System
QUEST) where a part of the computation was carried out.   We
acknowledge use of the HEALPIX package. We acknowledge the use of the
Legacy Archive for Microwave Background Data Analysis
(LAMBDA). Support for LAMBDA is provided by the NASA Office of Space
Science. 



\section*{References}
\begin{thebibliography}{99}

\bibitem{Komatsu:2011} E.~Komatsu, {\em et. al.},  ApJS, 192, 18 (2011)
\bibitem{Kim:2011} J.~Kim, C.~Park, G.~Rossi, S.~M.~Lee and
  J.~R.~Gott, JKAS, 44, 217 (2011)
\bibitem{Hwang:2012} J.-C.~Hwang, JKAS, 45, 65  (2012)

\bibitem{Toffolatti:1997dk} 
  L.~Toffolatti, F.~Argeso Gomez, G.~De Zotti, P.~Mazzei,
  A.~Franceschini, L.~Danese and C.~Burigana, 
  Mon.\ Not.\ Roy.\ Astron.\ Soc.\  {\bf 297}, 117 (1998)
  [astro-ph/9711085].
\bibitem{gold2011} B.~Gold, {\em et.al.}, ApJS, 192, 15 (2011).
\bibitem{lambdasite} http://lambda.gsfc.nasa.gov/
\bibitem{Scodeller:2012fi} 
  S.~Scodeller, F.~K.~Hansen and D.~Marinucci,
  Astrophys.\ J.\  {\bf 753}, 27 (2012)
  [arXiv:1201.5852 [astro-ph.CO]].
\bibitem{Scodeller:2012sw} 
  S.~Scodeller and F.~K.~Hansen,
  arXiv:1207.2315 [astro-ph.CO].
\bibitem{Chingangbam:2012} P.~Chingangbam and C.~Park, submitted to
  Journal of Physics Conference Series. 
\bibitem{Tomita:1986} H.~Tomita, Progr.~Theor.~Phys.~{\bf 76}, 952 (1986).
\bibitem{Coles:1988}  P.~Coles, Mon.\ Not.\ Roy.\ Astron.\ Soc.\  {\bf
  234}, 509 (1988).
\bibitem{Gott:1990} J.~R.~Gott, C.~Park, R.~Juzkiewicz, W.~E.~Bies,
  F.~R.~Bouchet and A.~Stebbins, Astrophys.~J. {\bf 352}, 1  (1990). 
\bibitem{Winitzki:1997jj}
  S.~Winitzki and A.~Kosowsky,
  New Astron.\  {\bf 3}, 75 (1998)
  [arXiv:astro-ph/9710164].
\bibitem{argueso} F.~Argueso, J.~Gonzalez-Nuevo and L.~Toffolatti,
  2003 ApJ {\bf 598} 86
\bibitem{boughn} S. P. Boughn, R. B. Partridge, 
2008, PASP {\bf 120}, No. 865
2008, Publications of the Astronomical Society of the Pacific {\bf 120}, No. 865
\bibitem{babich} D.~Babich and E.~Pierpaoli, 2008, Phys. Rev. D {\bf
  77},   123011 (2008)  
\bibitem{Lacasa:2011ej} 
  F.~Lacasa, N.~Aghanim, M.~Kunz and M.~Frommert,
  arXiv:1107.2251 [astro-ph.CO].
\bibitem{Munshi:2012kq} 
  D.~Munshi, P.~Coles and A.~Heavens,
  arXiv:1207.6217 [astro-ph.CO].
\bibitem{Schmalzing:1997uc} 
  J.~Schmalzing and K.~M.~Gorski,
  astro-ph/9710185.
\bibitem{Lim:2012} E.~A.~Lim and D.~Simon,
  J. Cosmol. Astropart. Phys., 1, 48 (2012)  
\bibitem{Hikage:2006fe}
  C.~Hikage, E.~Komatsu and T.~Matsubara,
  Astrophys.\ J.\  {\bf 653}, 11 (2006)
  [arXiv:astro-ph/0607284].
\bibitem{Hikage:2012bs} 
  C.~Hikage and T.~Matsubara,
  arXiv:1207.1183 [astro-ph.CO].
\bibitem{PLANCK} Planck Collaboration, Planck early results. VII. The
  Early Release Compact Source Catalogue, A\&A 536, A7 (2011)

\end{thebibliography}
\end{document}